\documentclass[a4paper]{jpconf}
\usepackage[utf8]{inputenc}
\usepackage{amsmath,amsfonts,amssymb,amscd,braket}
\usepackage{graphicx}
\usepackage{cite}
\usepackage{ytableau}
\usepackage[vcentermath]{youngtab}
\usepackage{tensor}

\newcommand{\SU}[1]{\ensuremath{\mathrm{SU}(#1)}}
\newcommand{\U}[1]{\ensuremath{\mathrm{U}(#1)}}

\newcommand*{\rep}[2][]{\ensuremath{{\boldsymbol{#2}#1}}} 

\renewcommand{\bar}[1]{\overline{#1}}

\newcommand{\inv}{\ensuremath{\mathcal{I}}}

\newcommand{\birdtrack}[2]{\parbox{#1}{\includegraphics[width=#1]{#2}}}

\newcommand{\tryoung}[1]{\raisebox{1.6pt}{\text{\,\tiny$\young(#1)$\,}}} 

\newcommand{\Eqref}[1]{equation~\eqref{#1}}

\newcommand{\Inv}[1]{\ensuremath{\mathcal{I}_{{#1}}}}
\newcommand{\Jnv}[1]{\ensuremath{\mathcal{J}_{{#1}}}}

\definecolor{Gray}{gray}{0.95}
\newcommand{\bbox}[1]{\fcolorbox{gray}{Gray}{~$\displaystyle #1$~}}

\allowdisplaybreaks[1]

\definecolor{dred}{HTML}{D95F02}
\colorlet{red}{white!15!dred}
\definecolor{darkgreen}{HTML}{1B9E77}
\definecolor{lightgray}{gray}{0.90}

\usepackage[bookmarks=false,hidelinks]{hyperref}

\ytableausetup{centertableaux,boxsize=1.2em} 
\ytableausetup{smalltableaux}

\begin{document}
\title{On the systematic construction of basis invariants}

\author{Andreas Trautner}

\address{Max-Planck-Institut f\"ur Kernphysik, Saupfercheckweg 1, 69117 Heidelberg, Germany}

\ead{trautner@mpi-hd.mpg.de}

\begin{abstract}
We describe a new, generally applicable strategy for the systematic construction of basis invariants (BIs).
Our method allows one to count the number of mutually independent BIs
and gives controlled access to the interrelations (syzygies) between mutually dependent BIs.
Due to the novel use of orthogonal hermitian projection operators, we obtain the shortest
possible invariants and their interrelations.
The substructure of non-linear BIs is fully resolved in terms of linear, basis-covariant objects.
The substructure distinguishes real (CP-even) and purely imaginary (CP-odd) BIs in a simple manner.
As an illustrative example, we construct the full ring of BIs of the scalar potential of the general Two-Higgs-Doublet model.
\end{abstract}


Everybody is used to the conventional way of setting up quantum field theory models:
One picks fields in certain representations of symmetries and the Lagrangian is parametrized
as linear combination of all symmetry invariant operators up to a certain dimension.
However, if there are multiple fields with exclusively the same quantum numbers, these fields 
are physically indistinguishable, implying that they may be mixed at will, without observable consequences. 
On the Lagrangian level, such a mixing of fields does, in fact, correspond to a 
mixing of symmetry invariant operators, thereby parametrizing the Lagrangian in different ways.
This arbitrariness in parametrization (or basis choice, in different words) 
must not affect physical statements derived from a model. 
Notwithstanding this, the presence of large basis change freedoms often obscures the physical 
properties of a model. 

In order to make the physical discussion as general and transparent as possible,
it seems worthwhile to use basis invariant (BI) objects.
An original arena for BI techniques was the detection of CP violation in the Standard Model (SM) \cite{Jarlskog:1985ht} and extensions 
\cite{Bernabeu:1986fc,Branco:1986gr,Botella:1994cs,Lavoura:1994fv}.
Here, a formulation in terms of basis invariants (BIs) immediately gets rid of spurious rephasings,
thereby allowing direct access to physical properties of the model.
Many more applications of BIs are conceivable and -- ultimately -- it should be possible to describe and relate
all physical observables, say $S$-matrix elements, correlation functions \textit{etc.}, in terms of BI objects.
Having such a formulation would be wonderful, but has to date not been put forward in general.

Here we solve a major technical problem which arises as the first step along the way to any BI formulation: 
Given a theory formulated in an arbitrary basis, how does one obtain basis independent quantities in a controlled manner?
Several different ways have been used to construct invariants in the literature 
(for a certainly incomplete list see e.g.\ \cite{Bernabeu:1986fc,Branco:1986gr,Botella:1994cs,Lavoura:1994fv, Branco:2005em, 
Davidson:2005cw, Haber:2006ue, Ivanov:2005hg, Nishi:2006tg,Dreiner:2007yz, Varzielas:2016zjc, Berger:2018dxg}),
none of which is entirely satisfactory for varying reasons. Several occurring shortcomings of previous approaches are 
at the same time advantages of our newly proposed method:
\begin{itemize}
\item It is completely clear for us when we can stop looking for new invariants, i.e.\ when we have found a complete set of \textit{independent} invariants.
\item The substructure of invariants is fully resolved in terms of basis-covariant \textit{building blocks} 
(which are \textit{linear} combinations of potential parameters that transform in irreducible representations under basis changes).
\item Relations between different invariants (syzygies) can be derived in a systematic way.
\item There are no intermediate steps where explicit basis choices are necessary.
\end{itemize}
\medskip

But what does \textit{independent} even mean in the context of basis invariants?
Clearly, any function of invariants is an invariant itself; mathematically speaking 
the invariants form a \textit{ring}. So before we describe the actual algorithm, 
let us introduce the jargon which will be used throughout:
\begin{itemize}
 \item \textit{Algebraic Independence} \\[0.1cm]
 An invariant $\inv_1$, is algebraically dependent on a set of invariants, say $\inv_{2,3,..}$, if and only if there is a polynomial $P$
 such that 
 \begin{equation}
 P\left(\inv_1,\inv_2,\inv_3,\dots\right)=0\;.
 \end{equation}
 If there is no such polynomial, the invariants are said to be algebraically independent.
 Whenever we say \textit{independent} invariants in this work, we mean \textit{algebraically independent}.
 \item \textit{Primary invariants} \\[0.1cm]
 The primary invariants are a maximal set of algebraically independent invariants of a given model 
 (such a set is never unique, but its size is).
 \item \textit{Generating set of invariants} \\[0.1cm]
  All invariants that \textit{cannot} be solved for in terms of a polynomial $P$ of other invariants, i.e.\
 \begin{equation*}
 \inv_i \neq P\left(\inv_j,\dots\right)\;,
 \end{equation*}
 belong to the generating set of invariants. Vice versa, having found a generating set, we can express 
 \textit{any} invariant of the model as a polynomial in the generating set of invariants,
 \begin{equation*}
 \inv=P\left(\inv_1,\inv_2\dots\right)\;.
 \end{equation*}
\end{itemize}
\medskip

Perhaps not surprisingly, the number of algebraically independent BIs is equal to the number of ``physical'' parameters of a model --
which are all parameters that remain after all possible basis changes have been used to absorb parameters.
Alongside the identification of a set of primary BIs, our method also determines the number of physical parameters of a model in an unambiguous way. 
Of course, for many models it is easier to determine this number ``by hand'', but it may not always be simple to do that.
\medskip

Let us now summarize our algorithm for the systematic construction of basis invariants. 
The invariants are constructed in three steps:
 \begin{itemize}
  \item[1.)] The construction of building blocks. These are \textit{linear combinations} of potential parameters that transform in irreducible representations (irreps)
  under basis changes.
  \item[2.)] Using the building blocks as input, one derives the so-called Hilbert series (HS) and (multigraded) Plethystic logarithm (PL) of the model.  
  The HS reveals the number and minimal order of the primary invariants, while the PL allows one to determine the number and order of invariants in the generating set of a ring.
  In addition, the multigraded PL also reveals the substructure of invariants, i.e.\ their decomposition in terms of the building blocks.
  \item[3.)] The explicit construction of the invariants from the building blocks.  
 \end{itemize}
The key to our new approach is a melange of techniques from group theory (used in step 1  and 3) and algebraic invariant theory (used for step 2).
Specifically, step 1 and 3 are done by projection with mutually orthogonal hermitian projection operators \cite{Keppeler:2013yla}.
These can conveniently be constructed from Young tableaux \cite{Alckock-Zeilinger:2016bss,Alcock-Zeilinger:2016cva,Alcock-Zeilinger:2016sxc,Keppeler:2017kwt} via birdtrack diagrams 
\cite{Cvitanovic:1976am,Cvitanovic:2008zz}.
For step 2 we rely on the powerful invariant theory functions HS and PL \cite{1994dg.ga.....8003G,Labastida:2001ts} 
(see \cite{Benvenuti:2006qr,Feng:2007ur,Jenkins:2009dy,Hanany:2010vu,Lehman:2015via,Henning:2015daa,Henning:2015alf,Henning:2017fpj} 
for introduction and applications in high energy physics).

By definition, there must exist relations between those BIs of the generating set which are not primary invariants.
The number and structure of these relations (in terms of the building blocks) can be inferred from
higher order terms of the multigraded PL. Having the BIs at hand,
a straightforward step 4.) is to construct these relations explicitly.

Most certainly, all of the above steps could be performed by other means.
For example, computing the composition of irreducible representations from tensor products, as 
in the first step, is, of course, a well studied exercise.
Not surprisingly then, the building blocks constructed in our Two-Higgs-Doublet-Model (2HDM) example below
coincide with vectors and tensors derived within the so-called bilinear approach 
\cite{Nagel:2004sw,Ivanov:2005hg,Maniatis:2006fs, Ivanov:2006yq, Maniatis:2007vn, Ivanov:2007de,Ferreira:2010hy, Ferreira:2010yh, Maniatis:2014oza, Ivanov:2014doa}.
More recently, also the HS of the 2HDM has been constructed in terms of these invariants \cite{Bednyakov:2018cmx}. 
Furthermore, many of the resulting BIs of the 2HDM (or combinations thereof) have been constructed by direct 
contraction of coupling tensors \cite{Davidson:2005cw,Gunion:2005ja}.
The HS and PL have also been used before to derive details on the ring of invariants 
of the SM and simple extensions \cite{Jenkins:2009dy,Hanany:2010vu}.

Our algorithm gains justification by the fact that it is the first to overcome, in one go, 
all of the above mentioned challenges.
Undoubtably, we gain great advantage from using orthogonal projectors 
in every step, especially in combination with the invariant theory functions. 
Presumably this leads to the shortest possible invariants \textit{in principle}.
This in turn also leads to the shortest possible relations among invariants. 
Finally, using the PL in the context of BIs allows us to obtain the number and structure of 
relations in a straightforward systematic way. 

\medskip
\paragraph{The ring of basis invariants of the 2HDM.}
Let us now outline our algorithm based on an example.
We illustrate the systematic construction of the full ring of BIs
for the 2HDM \cite{Trautner:2018ipq}.
In a basis independent manner, the scalar potential of the general 2HDM is parametrized as
\begin{equation}\label{eq:Potential}
 V=\Phi^\dagger_a\,Y\indices{^a_b}\,\Phi^b+\Phi^\dagger_a\,\Phi^\dagger_b\,Z\indices{^{ab}_{cd}}\,\Phi^c\,\Phi^d\;.
\end{equation}
Here $a,\dots,d=\left\{1,2\right\}$ denote identical Higgs field copies $\Phi^a$,
which transform as $(\rep2,1)$ under a $\SU2_{\mathrm{L}}\otimes\U1_{\mathrm Y}$ gauge symmetry.
Upper and lower indices distinguish fields transforming as 
$\rep{2}$ or $\rep{\bar{2}}$ under \SU2 Higgs-flavor basis changes.\footnote{%
Of course, $\rep{2}$ and $\rep{\bar{2}}$ are equivalent representations in
the technical sense. However, it makes much more sense for bookkeeping,
and also in hindsight of generalizations to \SU{N}, to treat them 
like conjugate representations here.}

Hermiticity and gauge invariance constrain the number of independent real parameters in the coupling tensors $Y$ and $Z$ to $4$ and $10$, respectively. 
Three more parameters could be eliminated by an explicit basis choice, leading to a total of $11$ ``physical'' parameters.
We don't do this here and keep working in a general basis.

Using standard methods, $Y$ and $Z$ decompose into irreducible basis-covariants as
\begin{align}\nonumber 
 Y~&\mathrel{\widehat=}~\rep{1}\oplus\rep{3}\;,\\
 Z~&\mathrel{\widehat=}~\rep{1}\oplus\rep{1}\oplus\rep{3}\oplus\rep{5}\;.
\end{align}
The covariants on the r.h.s. are \textit{linear} combinations of the potential parameters 
and will serve as our \textit{building blocks} for all further constructions.

Already at this step, we decide to use Young tableaux and their corresponding hermitian projection operators \cite{Keppeler:2013yla,Keppeler:2017kwt}
to perform this decomposition. This warrants that we keep covariants orthogonal to each other 
and exhibits their symmetry properties in the most explicit way.
The extraction of building blocks by projection then reads
\Yvcentermath2
\begin{align}\nonumber
 Y_{\rep{1}}&=\left[\tryoung{a,b}\right]\indices{_{a'b'}}\,Y^{a'b'},&
 &Z_{\rep{1}_{(1)}}=\left[\tryoung{ab,cd}\right]\indices{_{a'b'c'd'}}\,Z^{a'b'c'd'},& 
 &Z_{\rep{1}_{(2)}}=\left[\tryoung{ac,bd}\right]\indices{_{a'b'c'd'}}\,Z^{a'b'c'd'},& \\ \label{eq:buildingblocks}
 Y_{\rep{3}}^{ab}&=\left[\tryoung{ab}\right]^{a\;b}_{a'b'}\,Y^{a'b'},&
 &Z_{\rep{3}_{\phantom{(1)}}}^{bc}=\left[\tryoung{abc,d}\right]^{b\;c}_{a'b'c'd'}\,Z^{a'b'c'd'},& 
 &Z_{\rep{5}}^{abcd}=\left[\tryoung{abcd}\right]^{a\;b\;c\;d}_{a'b'c'd'}\,Z^{a'b'c'd'}.&\raisetag{25pt}
\end{align}
Here we denote by $\left[\dots\right]$ the hermitian projector corresponding to the enclosed tableaux. 
Recall that writing indices in a line in a Young tableaux means that they should be symmetrized while indices in columns
ought to be anti-symmetrized. However, our projectors are not always pure (anti-)symmetrizers:
to obtain hermitian and mutually orthogonal projectors a specific ordering of symmetrization and 
anti-symmetrization operations has to be obeyed \cite{Keppeler:2013yla,Alckock-Zeilinger:2016bss,Alcock-Zeilinger:2016cva,Alcock-Zeilinger:2016sxc}.
For example, the hermitian projection operator on the triplet representation reads
\begin{equation}\label{eq:YTP_example}
\left[\tryoung{abc,d}\right]~=~\frac32\;\birdtrack{10ex}{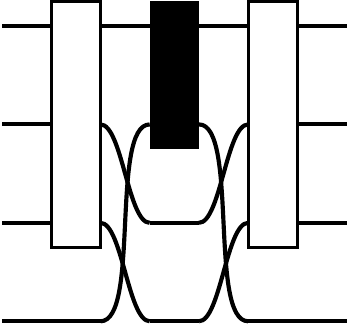}~=~\frac34\;\birdtrack{12.85ex}{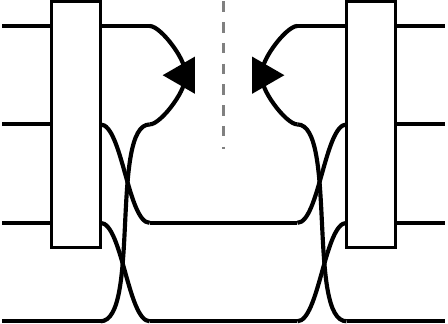}\;.
\end{equation}
On the r.h.s.\ are birdtrack diagrams which gives precise instructions on how to construct the corresponding projection operator, see \cite{Keppeler:2017kwt} for an introduction.
Focus on the middle of \Eqref{eq:YTP_example} for now. The open lines on the left and right correspond to upper and lower indices, respectively.
Connected lines means contraction of indices. 
The open(filled) boxes denote \mbox{(anti-)symmetrization} of the corresponding indices.
The fact that above operator reads the same from right-to-left and left-to-right
means that it is hermitian. Orthogonality of all operators is achieved if there 
are no crossed lines on the outermost legs of any projector.

All necessary projection operators for the 2HDM are explicitly stated in \cite{Trautner:2018ipq}.
Note that the \textit{effective} number of indices of the projection operators in \eqref{eq:buildingblocks} is often reduced. 
This is the case whenever the corresponding Young tableaux contains a complete column of [$n$ for $\SU{n}$] anti-symmetrized indices.
This can be understood as a (partly) \textit{factorization} of the corresponding operator \cite{Trautner:2018ipq}, illustrated on the r.h.s.\ of \eqref{eq:YTP_example}.
BIs naturally arise in this way whenever a projector fully factorizes to yield a ``bubble'' diagram.

We can directly represent the building blocks by symmetrized boxes.
The ``trivial'' linear invariants are given by 
\begin{equation}\label{eq:linearInvs}
 Y_{\rep{1}}~=~
 \begin{ytableau}
*(darkgreen) 1 \\
*(darkgreen) 2
\end{ytableau}\;, \quad
  Z_{\rep[_{(1)}]{1}}~=~
 \begin{ytableau}
*(lightgray) 1 & *(lightgray) 2 \\
*(lightgray) 3 & *(lightgray) 4
\end{ytableau}\;, \quad \text{and} \quad 
Z_{\rep[_{(2)}]{1}}~=~
 \begin{ytableau}
*(lightgray) 1 & *(lightgray) 3 \\
*(lightgray) 2 & *(lightgray) 4
\end{ytableau}\;.
\end{equation}
The non-trivially transforming building blocks read\footnote{%
We suppress indices in Young tableaux whenever they are ment to be assigned in the trivial way, i.e. 
incremental increase by one from left to right in each line.}
\ytableausetup{smalltableaux}
\begin{equation}
 Y_{\rep{3}}~\equiv~\ydiagram[*(darkgreen)]{2}\;,\quad Z_{\rep{3}}~\equiv~\ydiagram[*(red)]{2}\;,\quad Z_{\rep{5}}~\equiv~\ydiagram[*(lightgray)]{4}\;.
\end{equation}
We use a color coding here to denote indices from $Y$ (green) or $Z$ (red and gray).
For explicit expressions in terms of components of $Y$ and $Z$, see \cite{Trautner:2018ipq}.
\medskip

This completes the construction of building blocks. 
We will now use these non-trivially transforming building blocks to construct (the ring of) 
higher-order (non-linear in the potential parameters) BIs.
To find the number and structure of these invariants we make use of two 
very powerful functions of algebraic invariant theory: the Hilbert series
and Plethystic logarithm, see e.g.\ 
\cite{1994dg.ga.....8003G,Labastida:2001ts,Benvenuti:2006qr,Feng:2007ur,Jenkins:2009dy,Hanany:2010vu,Lehman:2015via,Henning:2015daa,Henning:2015alf,Henning:2017fpj}.
Knowing the representations of the building blocks we
assign a letter to each of them to ease the notation\footnote{%
$y$ stands for $Y$, while $t$ and $q$ are chosen to denote the \textit{t}riplet and \textit{q}uintuplet in $Z$, respectively.}
\begin{center}
 $\bbox{y~\mathrel{\widehat=}~Y_{\rep{3}}\;,\quad t~\mathrel{\widehat=}~Z_{\rep{3}}\;, \quad\text{and}\quad q~\mathrel{\widehat=}~Z_{\rep{5}}\;.}$
\end{center}
The Hilbert series computed from these representations (at $z=q=y=t$) is given by \cite{Bednyakov:2018cmx,Trautner:2018ipq}
\begin{equation}\label{eq:ungradedHS}
\mathfrak{h}(z)= \frac{1+z^3+4\,z^4+2\,z^5+4\,z^6+z^7+z^{10}}{\left(1-z^2\right)^4\left(1-z^3\right)^3\left(1-z^4\right)}\;.
\end{equation}
The multi-graded PL \cite{1994dg.ga.....8003G,Labastida:2001ts,Benvenuti:2006qr,Feng:2007ur} 
(expanded around zero for all arguments) is given by 
\begin{equation}\label{eq:PLgraded}
\begin{split}
 \mathrm{PL}\left[\mathfrak{H}\left(q,y,t\right)\right]=\,
 &t^2+t y+y^2+q^2+q t^2+q t y+q y^2+q^3+q t^2 y+q^2 t^2+q t y^2+q^2 t y \\
 &+q^2 y^2+q^2 t^2 y+q^2 t y^2 +q^3 t^3+q^3 t^2 y+q^3 t y^2+q^3 y^3-q^2 t^2 y^2 \\
 &-q^2 t^3 y^2-q^2 t^2 y^3-q^3 t^2 y^2-q^2 t^4 y^2-q^3 t^4 y-q^2 t^3 y^3-3 q^3 t^3 y^2 \\
 & -q^2 t^2 y^4-3 q^3 t^2 y^3-q^4 t^2 y^2-q^3 t y^4-\mathcal{O}\left(\left[tyq\right]^9\right)\;. \raisetag{20pt}
\end{split} 
\end{equation}
From the HS and PL we can simply read-off the following information:
\begin{itemize}
 \item The largest possible set of non-linear algebraically independent (i.e.\ primary) BIs 
 contains eight invariants; four of which are order 2, three of order 3 and one of order 4
 (corresponding to the factors in the denominator of \eqref{eq:ungradedHS}).
 \item The generating set of this ring contains $19$ invariants, and their orders and structures
 are given by the leading positive terms of the PL in \eqref{eq:PLgraded}.
\end{itemize}
Knowing their structure in terms of the building blocks (and more precisely, also the number of distinct invariants 
for each such configuration, which turns out to be always one here) we can go ahead and construct the invariants explicitly.

Again we will do this by projection. This time, the projectors will act on a tensor product of building blocks, 
projecting out from it the desired invariant. We denote tensor product BIs by their building block content.
An invariant with $a$ powers of $q$, $b$ powers of $y$, and $c$ powers of $t$ is denoted as
\begin{equation*}
 \Inv{a,b,c}
 ~\mathrel{\widehat=}~\left[Z_{\rep{5}}^{\otimes a}\otimes Y_{\rep{3}}^{\otimes b}\otimes Z_{\rep{3}}^{\otimes c}\right]_{\rep[_0]{1}}.
\end{equation*}
All required projection operators are very simple and always of the complete ``chocolate bar'' shape:\vspace{-0.5cm}
\ytableausetup{boxsize=1.5em}
\begin{equation}\label{eq:GenProjector}
\left[
\raisebox{1pt}{\scalebox{0.7}{
\begin{ytableau}
1 & 2 & \none[\cdots] &  n \\
\scriptstyle n+1 & \scriptstyle n+2 &  \none[\cdots] & 2n 
\end{ytableau}
}}
\right]
~=~
\birdtrack{35ex}{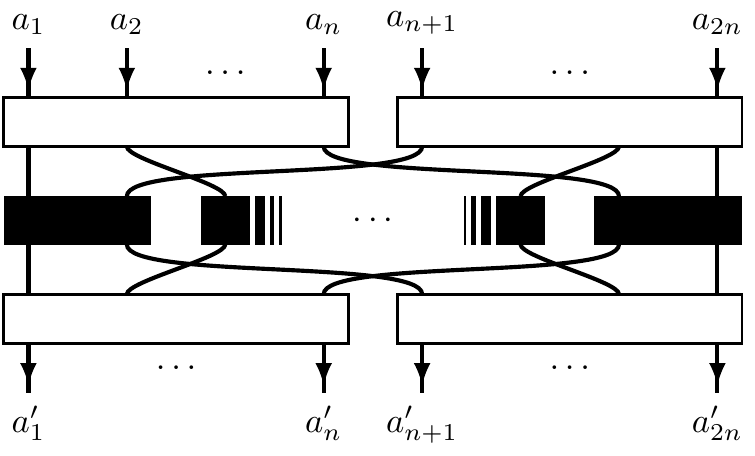}\;.\ytableausetup{smalltableaux}
\end{equation}
By inspection of \eqref{eq:ungradedHS} and \eqref{eq:PLgraded}, 
a possible choice for a maximal set of non-linear algebraically independent BIs then is given by\footnote{%
Most likely the choice of quartic invariant has to be modified for a Hironaka decomposition of the 2HDM, cf.\ the discussion in \cite[Sec.\ 8]{Trautner:2018ipq}.}
\begin{align}\notag
\Inv{2,0,0}:=&\ydiagram[*(lightgray)]{4,4}\;,\quad\,
\Inv{0,2,0}:=\ydiagram[*(darkgreen)]{2,2}\;,\quad\,
\Inv{0,1,1}:=\ydiagram[*(darkgreen)]{2}*[*(red)]{2,2}\;,\quad\,
\Inv{0,0,2}:=\ydiagram[*(red)]{2,2}\;,& \\ \label{eq:primaries}
\Inv{3,0,0}:=&\ydiagram[*(lightgray)]{6,6}\;, \qquad 
\Inv{1,2,0}:=\ydiagram[*(lightgray)]{4}*[*(darkgreen)]{4,4}\;, \qquad 
\Inv{1,0,2}:=\ydiagram[*(lightgray)]{4}*[*(red)]{4,4}\;, \quad \text{and}& \\\notag
 \Inv{2,1,1}:=&\ydiagram[*(lightgray)]{6,2}*[*(red)]{6,2+2}*[*(darkgreen)]{6,4+2}\;.&
\end{align}
Together with the linear invariants, \Eqref{eq:linearInvs}, these corresponds to a 
choice of $11$ algebraically independent BIs corresponding to the $11$ physical parameters of the 2HDM scalar sector.

In order to be able to express any given invariant as a polynomial of a fixed set of BIs,
one needs a generating set of BIs of a ring. The order and substructure of the
invariants of the generating set can be read-off from the PL, and their explicit form is, once again, 
obtained by projection. The resulting invariants read
\begin{align}\label{eq:secondaries}\notag
\Inv{1,1,1}&:=\ydiagram[*(lightgray)]{4}*[*(darkgreen)]{4,2}*[*(red)]{4,2+2}\;,&\\[5pt]\notag
\Inv{2,2,0}&:=\ydiagram[*(lightgray)]{6,2}*[*(darkgreen)]{6,6}\;,&             \Inv{2,0,2}&:=\ydiagram[*(lightgray)]{6,2}*[*(red)]{6,6}\;,& \\\notag
\Jnv{1,2,1}&:=\ydiagram[*(lightgray)]{4}*[*(darkgreen)]{5,3}*[*(red)]{5,5}\;,& \Jnv{1,1,2}&:=\ydiagram[*(lightgray)]{4}*[*(red)]{5,3}*[*(darkgreen)]{5,5}\;,& \\[5pt]\notag
\Jnv{2,2,1}&:=\ydiagram[*(lightgray)]{4}*[*(darkgreen)]{7,1}*[*(red)]{7,3}*[*(lightgray)]{7,7}\;,&
\Jnv{2,1,2}&:=\ydiagram[*(lightgray)]{4}*[*(red)]{7,1}*[*(darkgreen)]{7,3}*[*(lightgray)]{7,7}\;,& \\[5pt]\notag
\Jnv{3,3,0}&:=\ydiagram[*(lightgray)]{9,3}*[*(darkgreen)]{9,9}\;,&
\Jnv{3,0,3}&:=\ydiagram[*(lightgray)]{9,3}*[*(red)]{9,9}\;,& \\
\Jnv{3,2,1}&:=\ydiagram[*(lightgray)]{9,3}*[*(darkgreen)]{9,7}*[*(red)]{9,9}\;,& 
\Jnv{3,1,2}&:=\ydiagram[*(lightgray)]{9,3}*[*(red)]{9,7}*[*(darkgreen)]{9,9}\;.&
\end{align}
Explicit expressions for all of these invariants have been given in \cite[App D.]{Trautner:2018ipq}.
In the same work it was also shown that the only building blocks that transform non-trivially under CP 
are the triplets $Y_{\rep{3}}$ and $Z_{\rep{3}}$, which transform with a sign 
\begin{align}\notag
 &\mathcal{CP}\,:& &Y_{\rep{3}}^{ab}\mapsto -({Y_{\rep{3}}})_{ab}\;,&  &Z_{\rep[_{\phantom{(1)}}]{3}}^{ab}\mapsto -({Z_{\rep{3}}})_{ab}\;.& 
\end{align}
Therefore, BIs are CP odd if and only if they contain an odd total number of triplet building blocks.
We have denoted CP-odd BIs above by the letter $\mathcal{J}$ instead of $\mathcal{I}$.

Finally, we briefly outline the systematic construction of relations between BIs of a ring.
Empirically, each of the leading negative terms in the multi-graded PL, \Eqref{eq:PLgraded}, corresponds to a new independent relation
amongst the invariants \cite{Trautner:2018ipq}. The structure of the term corresponds to the structure of the relation, 
while its coefficient gives the number of independent relations of this structure.
To find the relation explicitly, we simply use an ansatz of suitable power products of lower-order BIs of the correct structure. 
The relation can then be obtained by solving a linear system \cite{Trautner:2018ipq}.

For example, using this strategy, one finds that the leading negative term in \eqref{eq:PLgraded}, $-q^2y^2t^2$, 
gives rise to the relation 
\begin{equation}\label{eq:Syz6}
\begin{split}
 3\,\Inv{1,1,1}^2~=&~2\,\Inv{2,1,1}\,\Inv{0,1,1} - \Inv{2,2,0}\,\Inv{0,0,2} - \Inv{2,0,2}\,\Inv{0,2,0} \\ 
 &+ 3\,\Inv{1,2,0}\,\Inv{1,0,2}+\Inv{2,0,0}\,\Inv{0,2,0}\,\Inv{0,0,2}-\Inv{2,0,0}\,\Inv{0,1,1}^2\;.
\end{split}
\end{equation}
Many more relations of this kind have been derived in \cite{Trautner:2018ipq}.
These previously mostly unknown relations can lead to important physical insights. 
For example, using them tremendously simplifies the derivation of necessary and sufficient conditions for 
CP conservation in the 2HDM \cite{Trautner:2018ipq}. 
Since it is possible to find a maximal set of algebraically independent invariants which are all CP-even, equations \eqref{eq:linearInvs} and \eqref{eq:primaries},
it must also be possible to express necessary and sufficient conditions for CP conservation 
solely in terms of CP-even quantities -- just as in the case of the SM, see e.g.\ \cite{Jenkins:2009dy}. Stating conditions in such a form may be beneficial,
since it allows one to distinguish between different, physically distinct, forms of CP conservation \cite{Ivanov:2018ime,Ivanov:2019kyh}.

\medskip
\paragraph{Final remarks.}
Further applications of our method are manifold. Using BIs and their interrelations,
we expect novel insights for the characterization of global symmetries and simplifications 
in the formulation of the renormalization group evolution \cite{Bednyakov:2018cmx, Bijnens:2018rqw}.
Immediate next steps are the application to the on-shell 2HDM and extension to the 3HDM. 

The method straightforwardly generalizes to other models. 
A bottleneck could become the construction of large hermitian projection operators, which scales with the number of to-be-symmetrized indices as $n!$ 
and quickly becomes very memory intensive. However, all of these projectors only have to be constructed \textit{once} -- 
a task which could be delegated to a super or perhaps quantum computer -- 
and then can be stored and reused with much less computational effort. 
Future aspirations should also target the formulation of physical observables
in terms of basis invariants, possibly using cuts of basis invariants and the amplitude formalism.

\section*{Acknowledgments}
The author wants to thank Igor P.\ Ivanov, Celso C.\ Nishi, and Jo\~ao P.\ Silva for discussions and motivation,
as well as Kevin Ingles for comments on the manuscript. 
This work has partly been supported by a postdoc fellowship of the German Academic Exchange Service (DAAD). 
The author gratefully acknowledges the hospitality of Stuart Raby and the Ohio State University during 
the manufacturing of this manuscript.


\bibliographystyle{iopart-num}
\section*{References}
\bibliography{Bibliography}

\end{document}